\PassOptionsToPackage{table}{xcolor}
\documentclass[sigconf]{acmart}

\usepackage{amsmath,amsfonts}
\usepackage{algorithm}
\usepackage{algorithmicx}
\usepackage{algpseudocode}
\usepackage{breqn}
\usepackage{graphicx}
\usepackage{subcaption} 
\usepackage{textcomp}
\usepackage{xcolor}
\usepackage{setspace}
\usepackage{booktabs}
\usepackage{xspace}
\usepackage{pifont}
\usepackage{enumitem}
\usepackage{multirow}
\usepackage{tabularx}
\usepackage{makecell} %
\newcolumntype{Y}{>{\centering\arraybackslash}X} %
\usepackage{appendix}
\usepackage{algpseudocode}
\usepackage{tikz}
\usetikzlibrary{decorations.pathreplacing,calc}

\algnewcommand{\Input}{\item[\textbf{Input:}]}
\algnewcommand{\Output}{\item[\textbf{Output:}]}

\newtheorem{definition}{Definition}
\newtheorem{proof}{Proof}

\newcommand{\squishlist}{
 \begin{list}{$\bullet$}
  { \setlength{\itemsep}{1pt}
   \setlength{\parsep}{1pt}
   \setlength{\topsep}{2.5pt}
   \setlength{\partopsep}{0.5pt}
   \setlength{\leftmargin}{1em}
   \setlength{\labelwidth}{1em}
   \setlength{\labelsep}{0.6em}
  }
 }
 \newcommand{\squishend}{
 \end{list}
}

\newcommand{\stitle}[1]{\vspace*{0.2em}\noindent{\bf #1\/}}
\newcommand{\method}{\textsc{SiriusDeliver}\xspace}

\algdef{SE}[SUBALG]{Indent}{EndIndent}{}{\algorithmicend\ }%
\algtext*{Indent}
\algtext*{EndIndent}

\setcopyright{none}
\begin{document}

\title{SiriusDeliver: Automating Data Warehouse Delivery at Tencent}

\author{
Haining Xie$^{\dagger}$,
Xiaokai Zhou$^{\ddagger}$,
Jiaming Yang$^{\dagger}$,
Siqi Shen$^{\dagger}$,
Ziwei Wang$^{\dagger}$, 
Yifeng Zheng$^{\dagger}$,
Tengyue Xu$^{\dagger}$, 
Yipeng Shi$^{\dagger}$,
Zefang Zong$^{\dagger}$,
Yang Li$^{\dagger}$,
Peng Chen$^{\dagger}$,
Jie Jiang$^{\dagger}$, 
Debiao He$^{\ddagger}$, 
Xiao Yan$^{\ddagger}$,
Jiawei Jiang$^{\ddagger}$ 
}

\affiliation{
  \institution{$^{\dagger}$TEG, Tencent Inc. \quad $^{\ddagger}$Wuhan University \\
$^{\dagger}$\texttt{%
\{hainingxie,besmingyang,siqishen,willziwang,yifengzheng,leooxu,
portershi,willzong,thomasyngli,pengchen,zeus\}@tencent.com} \\
$^{\ddagger}$\texttt{%
\{xiaokaizhou,hedebiao,yanxiaosunny,jiawei.jiang\}@whu.edu.cn}}
  \country{}
}

\renewcommand{\shortauthors}{Xie et al.}
\renewcommand{\authors}{Haining Xie, Xiaokai Zhou, Jiaming Yang, Siqi Shen, Ziwei Wang, Yifeng Zheng, Tengyue Xu, Yipeng Shi, Zefang Zong, Yang Li, Peng Chen, Jie Jiang, Debiao He, Xiao Yan, Jiawei Jiang}
\renewcommand{\thefootnote}{}









\begin{abstract}
Enterprise data warehouses (DWs) support business-critical analytics, but warehouse task delivery remains a complicated production process involving context retrieval, workflow configuration, code generation, platform submission, and failure diagnosis.
Although large language models (LLMs) and coding agents have improved software development, they are insufficient for production DW delivery, which requires dependency-aware orchestration, lifecycle-aware artifact control, and continuous adaptation to evolving platform practices.
We present \method, an end-to-end delivery automation agent for production warehouse task submission.
\method integrates three components: a hierarchical delivery agent that orchestrates warehouse skills, an artifact lifecycle control module that verifies and revises artifacts before and after platform execution, and a trace-driven skill evolution mechanism that maintains reusable skills from delivery trajectories.
We evaluate \method through offline datasets and large-scale production deployment on Tencent Cloud WeData.
Offline experiments on real-world warehouse delivery cases show that \method improves delivery success and automation efficiency over representative baselines.
During a two-month deployment across 6 business teams and 4 warehouse task types, \method served 3,600 monthly active users and supported 18,240 delivery sessions, achieving an 87.2\% end-to-end success rate and a 73.5\% autonomous submission rate.
A one-month A/B test shows that \method reduces median delivery time from 228 to 23 minutes and engineer effort from 95 to 11 minutes, while maintaining comparable final delivery success.

\end{abstract}

\maketitle
 
\section{Introduction}
\begin{figure}
    \centering
    \includegraphics[width=0.99\linewidth]{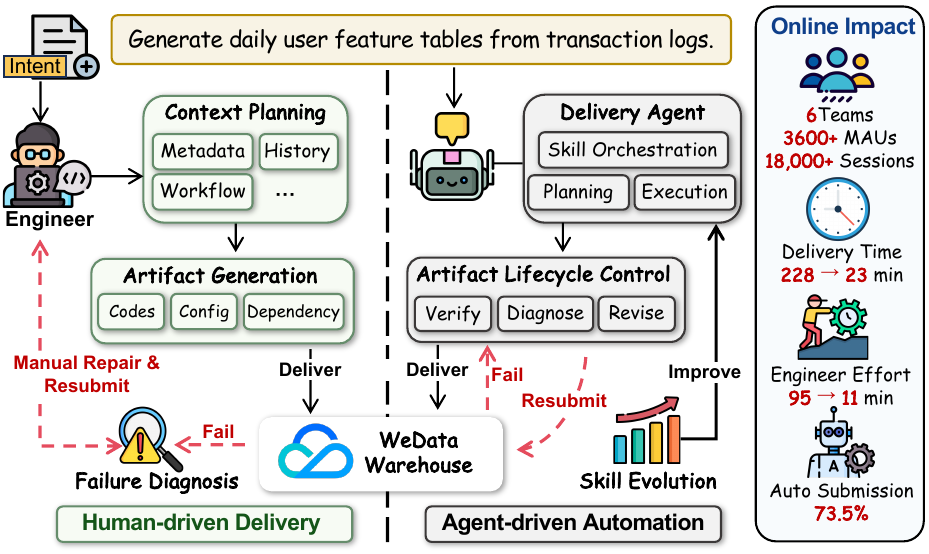}
\caption{Traditional human-driven warehouse delivery (left) vs. our agent-driven automation (right). \method automates the end-to-end delivery process, substantially reducing manual intervention in production warehouse systems.}
    \label{fig:motivation}
    \vspace{-3em}
\end{figure}

Enterprise data warehouses (DWs) underpin business-critical analytics by transforming raw business events into analytical tables.
At Tencent, the production warehouse platform serves over 10,000 monthly active data engineers, manages millions of warehouse tasks, and maintains hundreds of thousands of production tables.
Yet warehouse task delivery remains largely manual.
As shown in Figure~\ref{fig:motivation}, engineers must retrieve context from metadata services, workflow platforms, historical tasks, logs, and troubleshooting documents; generate code and configurations; validate dependencies; submit tasks; and repair failures.
This fragmented process is labor-intensive and error-prone: missing dependencies, inconsistent schedules, invalid permissions, or inefficient queries can delay delivery or introduce silent data-quality risks.
Evidence from our internal questionnaire survey, supplemented by in-depth follow-up interviews with experienced practitioners, suggests that SQL development accounts for less than 25\% of  end-to-end delivery time, while information retrieval and cross-platform coordination collectively constitute the primary sources of effort.


Recent advances in large language models (LLMs)~\cite{openai2024gpt4technicalreport,geminiteam2024gemini15unlockingmultimodal,anthropic2024claude35sonnet,deepseekai2026deepseekr1incentivizingreasoningcapability} and coding agents~\cite{codebuddy2026overview,openai2025codex,github2026copilot,anthropic2026claudecode} have demonstrated strong capabilities in automated software development.
However, production warehouse task delivery fundamentally differs from repository-level coding: it requires end-to-end coordination over data semantics, workflow dependencies, platform policies, execution feedback, and long-term operational maintenance.
Without such DataOps integration, general-purpose coding agents often produce plausible artifacts that remain incomplete, unsafe, or difficult to maintain in production warehouse environments.


Our production experience shows that directly applying general-purpose code agents to this setting leaves three key challenges.

\stitle{(1) Multi-stage task composition.}
Warehouse task delivery spans multiple interdependent stages—from metadata retrieval and dependency resolution to code generation, configuration, and platform submission.
For instance, importing operational data requires resolving source/target schemas, incremental keys, and schedule dependencies before task execution.
In practice, one-shot LLM planning frequently omits or misorders such prerequisites, producing invalid configurations or failed submissions.
Existing tool-use agents~\cite{schick2023toolformerlanguagemodelsteach,patil2023gorillalargelanguagemodel,qin2023toolllmfacilitatinglargelanguage} rely on curated supervision or static tool definitions, which fail to handle enterprise warehouses with complex, implicit, and team-specific dependency constraints.

\stitle{(2) Trustworthy artifact delivery.}
A deliverable warehouse bundle encompasses executable code, workflow topologies, schedules, permissions, and platform metadata.
Syntactically valid artifacts may still contain invalid dependencies, schedule conflicts, or logic bugs that trigger silent data errors or execution failures.
Existing code agents~\cite{yang2024sweagent,zhang2024autocoder} perform reactive post-execution repair based solely on logs.
In contrast, production delivery demands proactive pre-submission verification alongside post-execution diagnosis to guarantee end-to-end artifact safety.

\stitle{(3) Continuous platform adaptation.}
As platform APIs, data schemas, scheduling policies, and operational conventions continually evolve, static delivery rules can rapidly become obsolete.
Existing lifelong learning methods~\cite{shinn2023reflexionlanguageagentsverbal,wang2023voyageropenendedembodiedagent,zhao2024expelllmagentsexperiential} optimize transient task-level behaviors rather than auditable, reusable skill updates.
Furthermore, they overlook signals from successful trajectories, which can reveal potentially redundant context acquisition and reasoning steps that could streamline future execution.

To address these challenges, we present \method, an end-to-end agent for production warehouse task delivery.
For \textbf{Challenge (1)}, its \textit{delivery automation agent} iteratively plans, routes, executes, and updates memory over hierarchical scenario, context, artifact, and platform skills, composing their outputs into workflow specifications, executable artifacts, and task configurations.
For \textbf{Challenge (2)}, \textit{artifact lifecycle control} validates and revises artifacts before and after execution by combining deterministic platform evidence, LLM reasoning, execution feedback, and retrieved diagnostic knowledge to detect risks and explain failures.
For \textbf{Challenge (3)}, \textit{trace-driven skill evolution} groups trajectories by skill and outcome into structured evidence and applies bounded LLM updates, using failures for repair and successes for compression.

We evaluate \method through offline real-world cases and online production deployment.
\textbf{Offline}, on 200 real-world delivery cases across 4 warehouse scenarios, \method improves the average end-to-end delivery success rate by 14.5 points (from 71.5\% to 86.0\%) over the strongest skill-augmented baseline and reduces token consumption by 30\%, with ablations validating the contribution of each component.
\textbf{Online}, a two-month deployment on Tencent Cloud WeData across 6 business teams and 4 warehouse scenarios served 3,600 monthly active users and supported 18,240 delivery sessions, achieving an 87.2\% end-to-end success rate and a 73.5\% autonomous submission rate.
In a one-month A/B test, \method reduced delivery time from 228 to 23 minutes, engineer effort from 95 to 11 minutes, time to first artifact from 44 to 2.6 minutes, and manual intervention from 100\% to 21\%.


To summarize, we make the following contributions:
\squishlist
    \item We propose \method, a delivery automation agent that organizes warehouse skills hierarchically and generates platform-submittable artifact bundles from business requirements.

    \item We design artifact lifecycle control with pre- and post-execution diagnosis to detect risks, explain failures, and revise artifacts.

    \item We introduce trace-driven skill evolution, which turns delivery trajectories into bounded skill updates for failure repair and successful-run compression.

    \item We evaluate \method on 200 real-world cases and in large-scale production deployment, achieving an 87.2\% end-to-end success rate while reducing median delivery time and engineer effort by 89.9\% and 88.4\%, respectively.
\squishend


\section{Related Work}
\label{sec:related}

\stitle{Data warehouse platforms.}
Modern cloud-native data platforms (e.g., Snowflake~\citep{dageville2016snowflake}, BigQuery~\citep{melnik2010dremel}, and Delta Lake~\citep{armbrust2020delta}) provide high-performance storage and analytical query engines.
However, these systems primarily expose low-level engine capabilities rather than automating end-to-end task delivery.
Engineers must still manually interpret requirements, resolve dependencies, configure schedules, and diagnose execution failures. 
In contrast, \method operates as an automation layer above warehouse platforms to close the requirement-to-submission loop using execution feedback.

\stitle{LLM-based NL-to-SQL.}
NL-to-SQL maps natural language to executable queries via schema linking, in-context learning, and execution feedback~\citep{yu2018spider,pourreza2023din,gao2023dailsql,talaei2024chess}.
While effective for query synthesis, NL-to-SQL solves only a subset of production warehouse delivery. 
A deployable warehouse task requires not only valid SQL, but also workflow topology, scheduling parameters, permission configs, and cross-stage failure handling.
Thus, NL-to-SQL alone cannot satisfy full-lifecycle production submission requirements.

\stitle{LLM agents for code and data systems.}
LLM agents extend LLMs to repository editing and data analytics (e.g., SWE-agent~\citep{yang2024sweagent}, AutoCodeRover~\citep{zhang2024autocoder}, TaskWeaver~\citep{qiao2023taskweaver}, and Data Interpreter~\citep{hong2025data}), while platform assistants (e.g., Databricks Assistant~\citep{databricks2026assistant}, Snowflake Cortex~\citep{snowflake2026cortexanalyst}, and Gemini in BigQuery~\citep{google2026bigquerygemini}) assist in query generation.
However, these systems mainly focus on code completion or interactive analysis. 
They do not address production task delivery, which demands dependency-aware orchestration, lifecycle artifact verification before/after platform submission, and trace-driven skill maintenance as platform practices evolve.

\begin{figure}
    \centering
    \includegraphics[width=.99\linewidth]{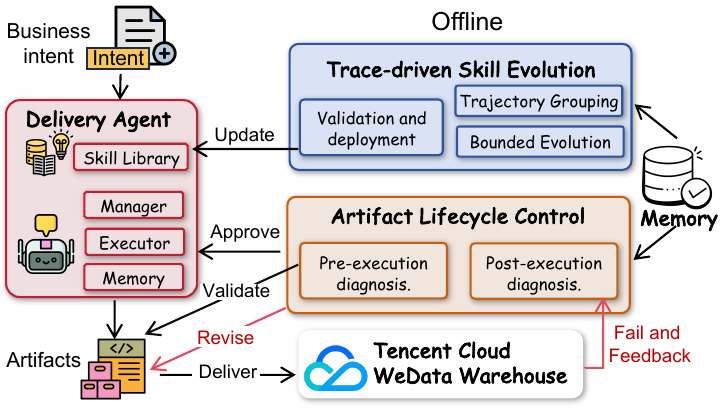}
    \caption{Overview of  key techniques in \method.}
    \label{fig:system-overview}
\end{figure}

\section{\method}

Figure~\ref{fig:system-overview} shows the workflow of \method.
Given a warehouse requirement, \method first invokes the \textit{delivery automation agent} to select warehouse skills, retrieve context, and generate an artifact bundle with workflow specifications, executable artifacts, and task configurations.
The bundle is then passed to \textit{artifact lifecycle control} for pre-execution diagnosis, where evidence, dependencies, permissions, and configurations are verified before submission.
Only approved artifacts are submitted to the warehouse platform by the agent.
After execution, post-execution diagnosis analyzes platform feedback and logs, explains failures, and revises the artifacts.
The revised artifacts are returned to the agent for resubmission, closing the delivery loop.
Throughout this process, \method records skill calls, intermediate artifacts, diagnostic results, execution feedback, and final outcomes as delivery trajectories.
Offline, \textit{trace-driven skill evolution} transforms these trajectories into structured evidence, and proposes bounded skill updates for failure repair and successful-run compression.

\subsection{Delivery Automation Agent}
Warehouse delivery starts from a business requirement, but executable artifacts depend on progressively acquired context, including data semantics, workflow states, historical configurations, and platform constraints.
A one-pass LLM generation process cannot reliably determine which evidence is missing, which skills should be invoked, or when the delivery state is complete.
General tool-use agents can invoke tools, but they usually do not explicitly track warehouse-specific prerequisites, dependencies, and submission readiness.
To address this, we propose a delivery automation agent that maintains a structured delivery state and orchestrates warehouse skills to incrementally construct a candidate artifact bundle.

\stitle{Delivery objective.}
The agent aims to produce a candidate artifact bundle for downstream lifecycle validation.
\begin{definition}[Artifact bundle]
Given a user requirement $q$, historical workflow configurations $H$, and platform context $M$, the delivery automation agent outputs
$
A=\langle G,C,P\rangle,
$
where $G$ is a workflow specification, $C$ is a set of node-level executable artifacts, and $P$ is a set of task configurations.
\end{definition}
$G$ describes the workflow structure, which can be a single node for single-task delivery or a DAG for workflow-level delivery.
$C$ contains executable artifacts such as SQL, PySpark, Flink SQL, or Python programs.
$P$ specifies schedules, dependencies, retry policies, runtime parameters, resource hints, permissions, and other platform options.
This allows code, workflow structure, and configurations to be jointly validated before platform submission.

\stitle{Design insights.}
Our design follows four observations from production warehouse delivery.
\textit{First,} evidence is scenario-specific: synchronization tasks require connections, schemas, mappings, and incremental keys, whereas computation tasks require schemas, partition semantics, dependencies, and execution resources.
\textit{Second,} submission-critical fields must be clarified rather than inferred.
\textit{Third,} existing warehouse assets should be reused whenever possible.
\textit{Fourth,} historical configurations provide grounded references for team-specific schedules, resources, retries, naming, and dependencies.
Accordingly, the agent identifies the delivery scenario, resolves critical fields, reuses existing workflows or nodes, and retrieves relevant configurations before artifact generation.

\stitle{Agent and warehouse skills.}
The delivery automation agent is a closed-loop framework that maintains the delivery state and orchestrates warehouse skills until all required artifacts are complete.

A warehouse skill is a reusable executable capability with defined inputs, outputs, prerequisites, and side effects, implemented through metadata retrieval, deterministic scripts, LLM calls, historical-task search, or platform APIs.
Besides task outputs, a skill may emit structured hints, such as the inferred scenario, unresolved prerequisites, and candidate next steps, to guide subsequent orchestration.
To reduce planning complexity, \method organizes warehouse skills into four conceptual layers.
\textit{Scenario skills} identify the delivery scenario, such as computation, synchronization, or workflow update.
\textit{Context skills} retrieve warehouse evidence, such as table metadata, workflow structure, and historical configurations.
\textit{Artifact skills} generate executable code, workflow specifications, and task configurations.
\textit{Platform skills} interact with the warehouse platform, including approved submission, status query, and feedback collection.
This hierarchy narrows the candidate skill space at each step and makes implicit dependencies easier to resolve.

\begin{figure}[!t]
    \centering
    \includegraphics[width=0.99\linewidth]{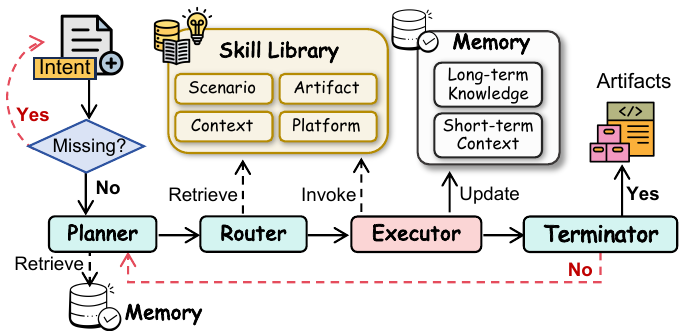}
    \caption{Orchestration loop of the delivery automation agent. The manager plans and routes warehouse skills, the executor invokes them and updates memory, and the loop iterates until a platform-submittable artifact bundle is produced.}
    \label{fig:delivery-agent}
\end{figure}

\stitle{Orchestration.}
Figure~\ref{fig:delivery-agent} illustrates the iterative delivery orchestration process.
Given a user requirement, the agent checks whether the current delivery state contains sufficient information for the next step.
It requests targeted clarification when critical inputs are missing; otherwise, it selects and executes warehouse skills based on the current state and historical context.
Skill outputs are normalized and written to memory, enabling subsequent steps to use newly acquired metadata, workflow evidence, generated artifacts, and platform feedback.
The loop terminates when $G$, $C$, and $P$ are complete, and the resulting candidate bundle is passed to artifact lifecycle control.
After lifecycle approval, the agent invokes the corresponding platform skill to submit the bundle and collect execution feedback.
The framework has three internal roles.
\squishlist
  \item \textit{Manager} tracks the delivery state and decides whether to request clarification, continue execution, or terminate with a complete candidate bundle.
  Its \emph{planner} decomposes tasks, \emph{router} selects skills, and \emph{terminator} checks bundle completeness.

  \item \textit{Executor} resolves inputs from the requirement and memory, schedules prerequisite-dependent skills, invokes them, and records normalized outputs and execution traces.

  \item \textit{Memory} stores short-term session context, including metadata, workflow evidence, and skill outputs, as well as long-term knowledge from compressed trajectories, reusable configurations, and execution patterns.
  It also retains platform feedback and diagnostic traces for subsequent reasoning, lifecycle control, and skill evolution.
\squishend

\subsection{Artifact Lifecycle Control}
\begin{figure}
    \centering
    \includegraphics[width=0.99\linewidth]{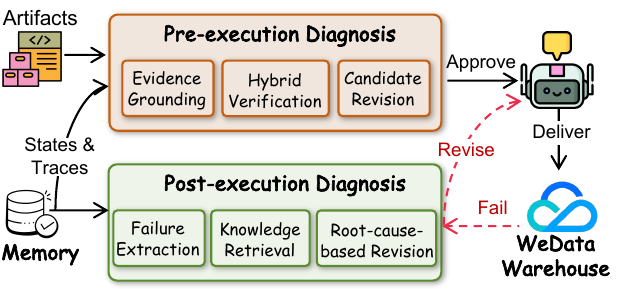}
\caption{Artifact lifecycle control. Artifacts undergo pre-execution verification before submission, and post-execution diagnosis transforms platform feedback into repair suggestions and revised artifacts for subsequent execution.}
    \label{fig:lifecycle-control}
    \vspace{-2em}
\end{figure}

The delivery agent produces candidate rather than production-ready artifact bundles.
Even syntactically valid bundles may contain silent semantic errors, invalid dependencies, permission or scheduling conflicts, and inefficient execution patterns.
Direct submission incurs costly trial and error, while pure LLM-based repair may overlook platform evidence, execution feedback, and historical diagnostic knowledge.
To address these risks, artifact lifecycle control validates and revises artifacts in two complementary stages, as shown in Figure~\ref{fig:lifecycle-control}: \textit{pre-execution diagnosis} verifies artifacts before submission using platform evidence, deterministic checks, and metadata-aware reasoning; \textit{post-execution diagnosis} analyzes platform feedback, retrieves diagnostic knowledge, and revises failed artifacts for subsequent execution.


\stitle{Diagnosis coverage.}
Based on practical observations from warehouse development, we define diagnosis coverage for both pre-submission risks and post-execution failures, as summarized in Table~\ref{tab:diag_coverage}.
Pre-execution diagnosis targets defects that can be detected before platform submission from generated artifacts, task configurations, metadata, and pre-compilation evidence.
For example, it detects brute-force scans caused by missing partition predicates and business-intent mismatches that may produce silent data errors.
Post-execution diagnosis targets failures exposed by platform feedback after compilation, dry-run, or execution.
For example, it diagnoses permission exceptions from access-denied logs and table-structure exceptions from partition conflicts.

A \emph{diagnosis item} is the basic unit of this coverage.
Each item specifies:
(1) the issue description and its risk;
(2) judgment conditions for reporting the issue;
(3) required evidence, such as code locations, compiler outputs, metadata, logs, or retrieved cases; and
(4) the execution workflow for collecting evidence and producing the diagnosis.
This structure prevents unsupported or speculative reports and makes diagnosis results comparable across tasks.

\stitle{Pre-execution diagnosis.}
To detect both execution failures and silent warehouse defects before costly production platform runs, we adopt a hybrid strategy that combines deterministic platform evidence with LLM-based reasoning.
Deterministic evidence grounds each diagnosis in concrete, directly verifiable platform facts, such as current metadata, task configurations, declared dependencies, and pre-compilation results.
LLM-based reasoning complements these facts by analyzing user intent, SQL semantics, and complex metadata-dependent risks that are difficult to reliably capture with rules alone.
The resulting diagnostic process contains three steps.


\squishlist
    \item \textit{Evidence grounding.}
    We construct a canonical diagnosis input from codes, configurations, user requests, and conversation history.
    Missing parameters are inferred only when supported by existing evidence, following the priority of code and configurations, conversation history, and code comments.
    If required fields remain unresolved, we ask targeted follow-up questions.

    \item \textit{Hybrid verification.}
      We combine statement-level pre-compilation with metadata-aware reasoning.
      The process decomposes SQL into atomic statements, collects pre-compilation evidence, extracts table and field references, retrieves related metadata, and checks predefined diagnosis items.

    \item \textit{Candidate revision.}
    We report only defects that may cause execution failures, data errors, or severe performance issues.
Each issue includes its type, severity, code evidence, and repair suggestion.
When sufficient evidence is available, we prompt an LLM with the original artifact, verified evidence, and repair suggestion to generate a revised artifact.
Otherwise, we preserve the original artifact and report unresolved issues.
\squishend

\stitle{Post-execution diagnosis.}
To convert platform failures into structured repair evidence, we build on log-grounded diagnosis~\cite{shen2026siriushelper, xu2025logsage, chen2024raccopilot} by extracting failure signals from platform feedback and retrieving relevant diagnostic knowledge to guide artifact revision.
This avoids direct repair from raw logs, reduces unsupported root causes, and keeps repair suggestions traceable.
It contains three steps.
\squishlist
    \item \textit{Failure extraction.}
    We collect detailed execution logs, runtime errors, task states, and related platform metadata.
    We then extract concise, structured failure signals from platform feedback, such as the exception type, failed operator, error message, task stage, or platform component.
    These signals must be directly derived from platform feedback rather than inferred from the SQL alone.

    \item \textit{Knowledge retrieval.}
    We retrieve relevant cases, rules, and troubleshooting documents from a diagnostic knowledge base using the extracted failure signals.
    The retrieved evidence provides historical fixes and known platform behaviors that are difficult to infer from the current failure alone.

    \item \textit{Root-cause-based revision.}
    We diagnose the failure using the original artifact, extracted failure signals, and retrieved knowledge.
    Raw logs are used as auxiliary context, while localization focuses on the key failure signals.
    The output includes an exception category, a concise root-cause explanation, and an actionable repair suggestion.
    When the repair is sufficiently grounded, we prompt an LLM with the original artifact, localized evidence, and repair suggestion to generate a revised artifact for the next execution.
    
\squishend

\begin{figure}
    \centering
    \includegraphics[width=0.99\linewidth]{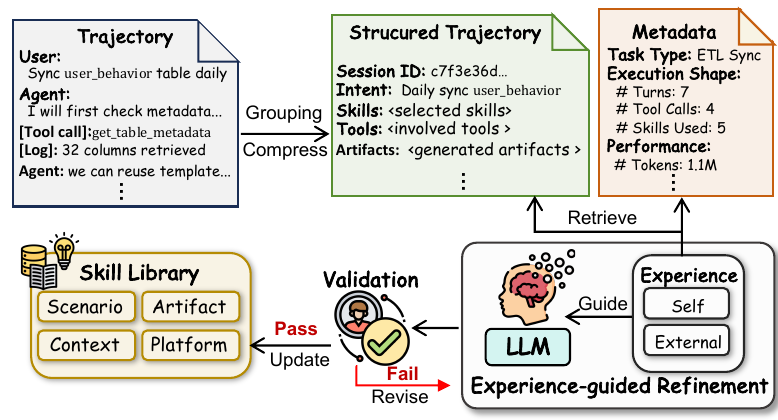}
    \caption{Trace-driven skill evolution. Raw trajectories are transformed into structured experience and metadata, which guide LLM-based skill refinement until validated updates are incorporated into the skill library.}
    \label{fig:skill-evolution}
    \vspace{-1.5em}
\end{figure}

\subsection{Trace-driven Skill Evolution}
\label{sec:failure-to-evolution}
Warehouse skills cannot remain static after deployment, because platform conventions, task patterns, and failure cases continuously change in production.
Online delivery further produces reusable experience, including successful execution paths, failed attempts, diagnostic results, and platform feedback.
However, these signals are scattered across long trajectories and cannot directly improve reusable warehouse skills.
Manual maintenance is costly, while unrestricted LLM rewriting may introduce unstable changes to operational skills.
To address this, we propose trace-driven skill evolution, an offline mechanism that converts delivery trajectories into bounded and reviewable skill refinements.
As shown in Figure~\ref{fig:skill-evolution}, this module proceeds in three steps. 
\textit{Evidence preparation} groups and compresses raw trajectories into structured evidence. \textit{Experience-guided refinement} proposes bounded updates from trajectory evidence and historical update records.
\textit{Validation and deployment} verify candidate skill updates before human-reviewed release and rollback-enabled production deployment.

\stitle{Design insights.}
Our design follows three observations from production deployment.
\textit{First}, failures often recur around specific skills, scenarios, or platform constraints and should therefore be aggregated at the skill level.
\textit{Second}, successful trajectories also reveal reusable configurations, stable execution patterns, and redundant context acquisition.
\textit{Third}, unrestricted LLM rewriting may propagate regressions across downstream submissions.
Therefore, \method derives structured evidence from both outcomes, applies explicit, bounded refinements, and releases updates only after offline replay validation and mandatory human review.

\stitle{Evidence preparation.}
We convert long, multi-skill delivery sessions into structured evidence for skill-level maintenance.
Each trajectory records the request, invoked skills and tools, intermediate artifacts, diagnoses, platform feedback, and final outcome, and may contribute to multiple skill groups.
We group trajectories by skill and outcome: failures reveal missing preconditions, unsafe assumptions, and weak recovery or validation, whereas successes reveal reusable context, shorter execution paths, and redundant calls.
For each selected session, we retain these key interactions in a compact trajectory and pair it with metadata on turns, skills, tool calls, latency, and errors, while removing verbose reasoning and truncating long outputs.
This representation preserves the evidence needed for evolution with lower noise and processing cost.



\stitle{Experience-guided refinement.}
This stage converts prepared evidence into bounded skill refinements.
Directly rewriting a skill from raw traces is unstable, as the LLM may overfit to one failure or change unrelated behavior.
We therefore retrieve two types of evolution experience before prompting the LLM.
\textit{Self-guided experience} retrieves previous updates of the same skill and their modification reasons, helping avoid repeated ineffective changes and maintain revision consistency.
\textit{External-guided experience} selectively retrieves relevant update records from related skills, providing transferable cross-skill repair or compression patterns.

Given the structured trajectories and retrieved experience, the LLM proposes a candidate refinement with an update reason linked to supporting evidence.
Repair-oriented refinements address recurring failures, while compression-oriented refinements remove redundant context, steps, or tool calls from successful executions.
An update budget strictly limits the number of modified fields and rewritten text in each iteration, keeping the evolution process reviewable and preventing free-form skill rewrites.

\stitle{Validation and deployment.}
This stage prevents skill evolution from introducing regressions into production delivery.
Candidate updates are not deployed immediately.
We first run offline validation on historical trajectories.
The validation set includes targeted failed trajectories, successful trajectories that should remain unaffected, and held-out trajectories from related skills.
A refinement is accepted only if it resolves the targeted failure or efficiency issue without introducing regressions on previously successful cases.

After offline validation, the candidate refinement and its update reason are sent for human review.
The reviewer may accept, reject, or edit the refinement before deployment.
For each accepted update, we record the previous skill version, the new version, the update reason, and the validation result.
If later monitoring detects degraded performance, the deployed skill can be rolled back to the previous version.
This protocol makes production skill evolution trace-driven, offline replay-validated, and safely reversible.

\begin{table*}[t]
\centering
\small
\caption{Offline end-to-end delivery performance and ablation on 200 WeData cases (50 per task type).
The upper block compares three coding products under two configurations (\textit{+ skills} vs.\ \textit{+ \method}); $\Delta$ is the average-success gain of \textit{+ \method}.
The lower block ablates the three \method components on the best product (Claude Code).
Success is mean $\pm$ 95\% CI ($n{=}200$); turns, time, and tokens are means.
\textbf{Bold}: best; \underline{underline}: second-best.}
\vspace{-1em}
\label{tab:offline-main}
\resizebox{\textwidth}{!}{
\begin{tabular}{llcccccccc}
\toprule
\multirow{2}{*}{Product} & \multirow{2}{*}{Configuration}
& \multicolumn{4}{c}{End-to-end success rate (\%) $\uparrow$}
& \multirow{2}{*}{Avg. success $\uparrow$}
& \multirow{2}{*}{Turns $\downarrow$}
& \multirow{2}{*}{Time (min) $\downarrow$}
& \multirow{2}{*}{Tokens (K) $\downarrow$} \\
\cmidrule(lr){3-6}
& & RT Sync & Off. Sync & RT Comp. & Off. Comp. & & & & \\
\midrule
\multirow{3}{*}{\textit{Codex}}
& + skills
& 72.0 & 68.0 & 62.0 & 56.0
& 64.5 $\pm$ 6.6
& 9.8
& 19.4
& 151.6 \\
& + \method
& 88.0 & 86.0 & 80.0 & 74.0
& 82.0 $\pm$ 5.3
& 6.1
& 11.8
& 98.4 \\
\rowcolor{gray!8}
& $\Delta$
& +16.0 & +18.0 & +18.0 & +18.0
& +17.5 & $-$3.7 & $-$7.6 & $-$53.2 \\
\midrule
\multirow{3}{*}{\textit{CodeBuddy}}
& + skills
& 76.0 & 72.0 & 66.0 & 60.0
& 68.5 $\pm$ 6.4
& 9.1
& 17.6
& 137.2 \\
& + \method
& 90.0 & 88.0 & 82.0 & 76.0
& \underline{84.0 $\pm$ 5.1}
& \underline{5.7}
& \underline{10.8}
& \underline{91.5} \\
\rowcolor{gray!8}
& $\Delta$
& +14.0 & +16.0 & +16.0 & +16.0
& +15.5 & $-$3.4 & $-$6.8 & $-$45.7 \\
\midrule
\multirow{3}{*}{\textit{Claude Code}}
& + skills
& 80.0 & 74.0 & 68.0 & 64.0
& 71.5 $\pm$ 6.2
& 8.4
& 16.0
& 124.0 \\
& + \method
& \textbf{92.0} & \textbf{90.0} & \textbf{84.0} & \textbf{78.0}
& \textbf{86.0 $\pm$ 4.8}
& \textbf{5.4}
& \textbf{10.2}
& \textbf{86.9} \\
\rowcolor{gray!8}
& $\Delta$
& +12.0 & +16.0 & +16.0 & +14.0
& +14.5 & $-$3.0 & $-$5.8 & $-$37.1 \\
\midrule
\multicolumn{10}{l}{\textit{Ablation of \method (on Claude Code)}} \\
\multicolumn{2}{l}{\quad w/o hierarchical skill orch.}
& 72.0 & 72.0 & 80.0 & 74.0
& 74.5 $\pm$ 6.0
& 7.7
& 14.8
& 118.7 \\
\multicolumn{2}{l}{\quad w/o artifact lifecycle control}
& 86.0 & 84.0 & 74.0 & 68.0
& 78.0 $\pm$ 5.7
& 7.1
& 13.7
& 111.8 \\
\multicolumn{2}{l}{\quad \quad w/o pre-execution diag.}
& 88.0 & 86.0 & 76.0 & 70.0
& 80.0 $\pm$ 5.5
& 6.7
& 12.9
& 101.4 \\
\multicolumn{2}{l}{\quad \quad w/o post-execution diag.}
& 88.0 & 86.0 & 78.0 & 72.0
& 81.0 $\pm$ 5.4
& 5.9
& 11.4
& 107.8 \\
\multicolumn{2}{l}{\quad w/o trace-driven skill evolution}
& 90.0 & 88.0 & 82.0 & 76.0
& 84.0 $\pm$ 5.1
& 6.2
& 11.7
& 112.3 \\
\bottomrule
\end{tabular}
}
\vspace{-1.5em}
\end{table*}

\section{Experimental Evaluation}
We evaluate \method through offline benchmarks and online production deployment by examining three questions:
\ding{172} whether it improves end-to-end warehouse delivery over representative baselines;
\ding{173} how artifact lifecycle control and trace-driven skill evolution affect reliability and efficiency; and
\ding{174} whether these gains translate into practical production benefits.



\subsection{Experimental Settings}
\label{sec:exp-settings}

\stitle{Dataset.}
We evaluate \method in both offline and online settings.
For offline evaluation, we construct a frozen benchmark of 200 production delivery cases from Tencent Cloud WeData, stratified equally across four task types (50 each): real-time synchronization, offline synchronization, real-time computation, and offline computation.
Each case pairs a user requirement with the platform context needed for execution.
The benchmark snapshots are frozen before evaluation, and all skill-evolution trajectories are strictly disjoint from evaluation cases to prevent data leakage (details in Appendix~\ref{app:benchmark}).
Due to proprietary schemas and platform metadata, these cases cannot be released; we instead provide DataClawEval\footnote{\url{https://github.com/Dicemy/DataClawEval}}, a public companion benchmark covering general data-engineering capabilities.
The online evaluation uses long-running real-world production deployment logs, as detailed in Section~\ref{sec:online_performance}.

\stitle{Baselines and product configurations.}
We compare three representative agentic coding products: CodeBuddy~\cite{codebuddy2026overview}, Claude Code~\cite{anthropic2026claudecode}, and Codex~\citep{openai2025codex}.
Unless otherwise specified, they use their native default backbones: GLM-5.2, Claude Opus 4.8, and GPT-5.5, respectively.
For each product, \textit{+ skills} adds the same base warehouse skill library, including identical skill interfaces, input schemas, executable implementations, and platform tools; \textit{+ \method} additionally enables hierarchical skill orchestration, artifact lifecycle control, and trace-driven skill evolution.
Within each product pair, the backbone, product version, skill library, platform tools, execution environment, and refinement budget are fixed, so the paired difference measures the incremental benefit of \method.
Because products retain different native backbones and internal agent implementations, their absolute results represent product-conditioned systems rather than isolated comparisons of agent frameworks or backbone capabilities.

\begin{table*}[t]
\centering
\scriptsize
\setlength{\tabcolsep}{4pt}
\renewcommand{\arraystretch}{1.08}
\caption{Paired evaluation of \method across representative LLM backbones on the same 200 cases (50 per task type), all run under the same CodeBuddy runtime.
$\Delta$ Success is the absolute gain of \textit{+ \method} over \textit{+ Skills}.}
\vspace{-2em}
\label{tab:paired-backbones}

\resizebox{\textwidth}{!}{
\begin{tabular}{llcccccc}
\toprule
Backbone
& Configuration
& Success (\%) $\uparrow$
& $\Delta$ Success (pp) $\uparrow$
& First-pass (\%) $\uparrow$
& Turns $\downarrow$
& Time (min) $\downarrow$
& Tokens (K) $\downarrow$ \\
\midrule
DeepSeek V4 Flash
& Agent + Skills
& 58.0 & -- & 42.0 & 11.2 & 21.8 & 162.0 \\
\rowcolor{gray!8}
& Agent + \method
& 74.0 & \textbf{+16.0} & 56.0 & 7.6 & 14.6 & 116.0 \\
\addlinespace[2pt]
GPT-5.5
& Agent + Skills
& 68.0 & -- & 54.0 & 9.2 & 17.6 & 130.0 \\
\rowcolor{gray!8}
& Agent + \method
& 80.0 & +12.0 & 64.0 & 6.4 & 12.2 & 100.0 \\
\addlinespace[2pt]
Claude Opus 4.8
& Agent + Skills
& 70.0 & -- & 55.0 & 9.0 & 17.4 & 132.0 \\
\rowcolor{gray!8}
& Agent + \method
& \textbf{84.0} & +14.0 & \textbf{68.0} & 6.0 & 11.4 & 95.0 \\
\addlinespace[2pt]
GLM-5.2
& Agent + Skills
& 68.5 & -- & 53.0 & 9.1 & 17.6 & 137.2 \\
\rowcolor{gray!8}
& Agent + \method
& \textbf{84.0} & +15.5 & \textbf{68.0} & \textbf{5.7} & \textbf{10.8} & \textbf{91.5} \\
\bottomrule
\end{tabular}
}
\vspace{-1.5em}
\end{table*}

\stitle{Metrics.}
We use five metrics.
\textit{End-to-end delivery success rate} is the fraction of cases producing platform-accepted and executable warehouse tasks whose artifacts, dependencies, and configurations pass manual verification.
\textit{First-pass success rate} is the fraction completed by the initial submission without refinement.
\textit{Turns} counts agent reasoning and action rounds; \textit{Time} measures wall-clock duration from receiving the requirement to the final submission result; and \textit{Tokens} counts total LLM input and output tokens.
For each product or backbone pair, we report the absolute percentage-point success improvement after enabling \method.
Because products may use different tokenizers and internal inference procedures, turns, time, and tokens are interpreted primarily within matched pairs rather than by their absolute values across products.

\stitle{Implementation and controls.}
Each case permits one initial submission and at most two refinement attempts based on platform feedback.
For fair paired comparisons, configurations share task inputs, platform context, permissions, skill access, and refinement budgets.
Agent states and temporary execution contexts are reset between cases to prevent cross-case information carryover.
To prevent leakage, evolution trajectories are disjoint from evaluation cases and the resulting skill snapshot is frozen before evaluation.
The backbone study uses the same 200 cases and CodeBuddy runtime, varying only the underlying model.
Additional isolation and backbone controls are provided in Appendix~\ref{app:evaluation-controls}.

\subsection{Offline Evaluation}
\label{sec:exp-offline}

\stitle{End-to-end performance.}
Table~\ref{tab:offline-main} reports the end-to-end delivery performance on 200 offline cases.
\method consistently outperforms all code-agent baselines across the four task types in both delivery effectiveness and efficiency.
Compared with the best baseline, \method improves the average end-to-end delivery success rate by 14.5 points (from 71.5\% to 86.0\%), with the largest per-type improvement of 16.0 points on offline synchronization and real-time computation tasks.
Meanwhile, \method reduces the average number of turns from 8.4 to 5.4, shortens the end-to-end delivery time from 16.0 to 10.2 minutes, and decreases token consumption by 30\% (from 124.0K to 86.9K).
These results show that \method improves reliability without relying on larger reasoning budgets.
The gain comes from structured skill orchestration, which reduces missing-context and wrong-skill-selection errors, and artifact lifecycle control, which prevents invalid artifacts from being repeatedly submitted and repaired through costly trial and error.
Among the four task types, offline computation obtains the lowest success rate because it often involves complex SQL or PySpark semantics, multi-table dependencies, historical workflow reuse, partition management, and scheduling configurations.
Nevertheless, \method still achieves a success rate of 78.0\% on challenging offline computation and maintains consistently lower overall turns, time, and token cost than the baselines, showing its effectiveness in the most difficult offline delivery setting.

\stitle{Effect of artifact lifecycle control.}
We evaluate artifact lifecycle control at both the pre-execution and
post-execution stages.
Pre-execution diagnosis detects 75.9\% of real artifact issues with a
false-positive rate of 9.5\%, while introducing an average latency of
only 6.1\,s.
Post-execution diagnosis directly repairs 73.0\% of failed executions
within one diagnosis-guided revision, with an average latency of
24.2\,s.
Together with the component ablations in
Table~\ref{tab:offline-main}, these results show that lifecycle control
Artifact lifecycle control improves delivery reliability by blocking risky artifacts before submission and converting a large fraction of execution failures into successful repairs, with modest runtime overhead.

\begin{figure}[!t]
  \centering
  \begin{subfigure}[t]{0.54\linewidth}
    \centering
    \includegraphics[width=\linewidth]{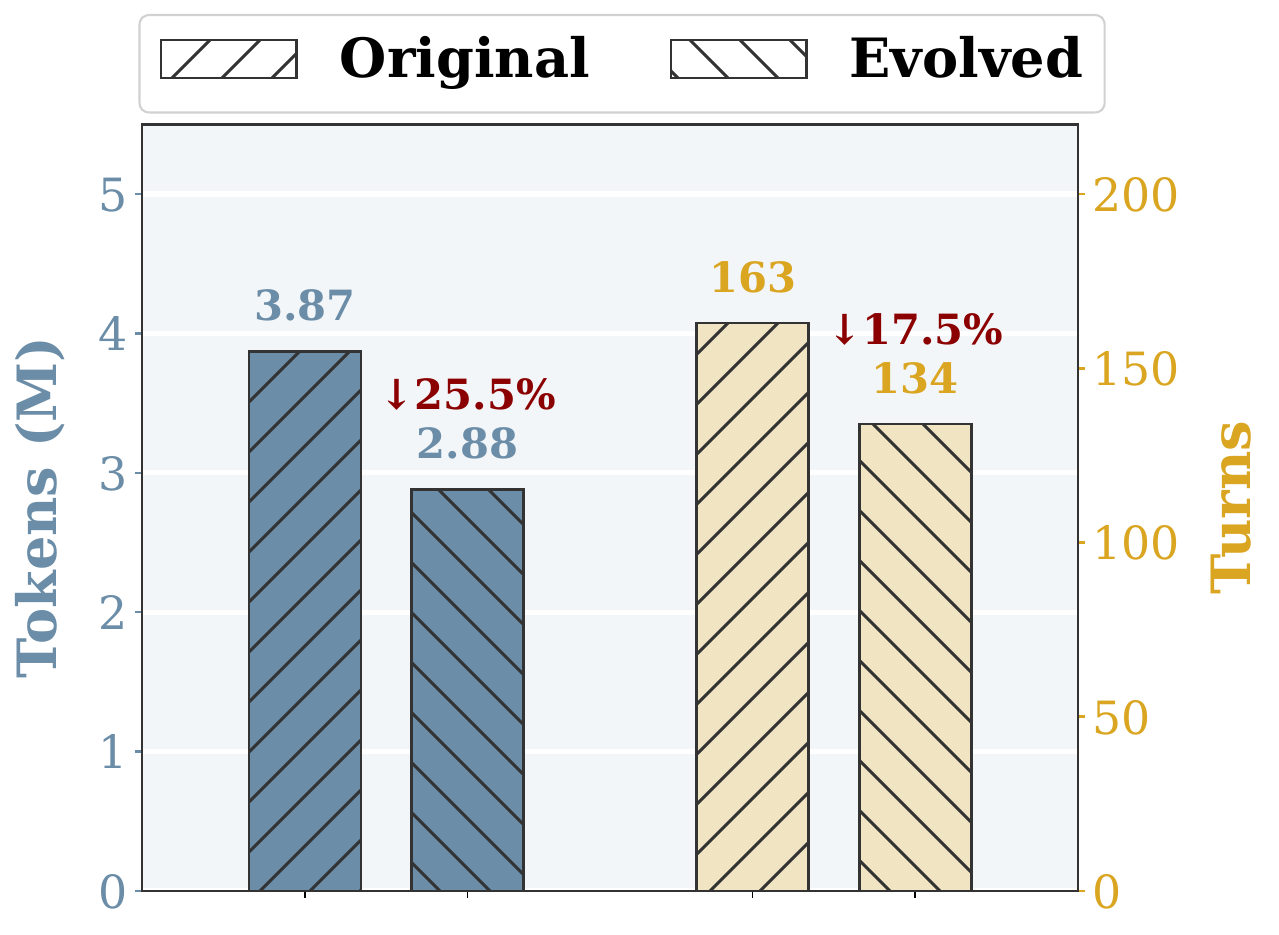}
    \label{fig:evolved_tokens}
  \end{subfigure}
  \hfill
  \begin{subfigure}[t]{0.45\linewidth}
    \centering
    \includegraphics[width=\linewidth]{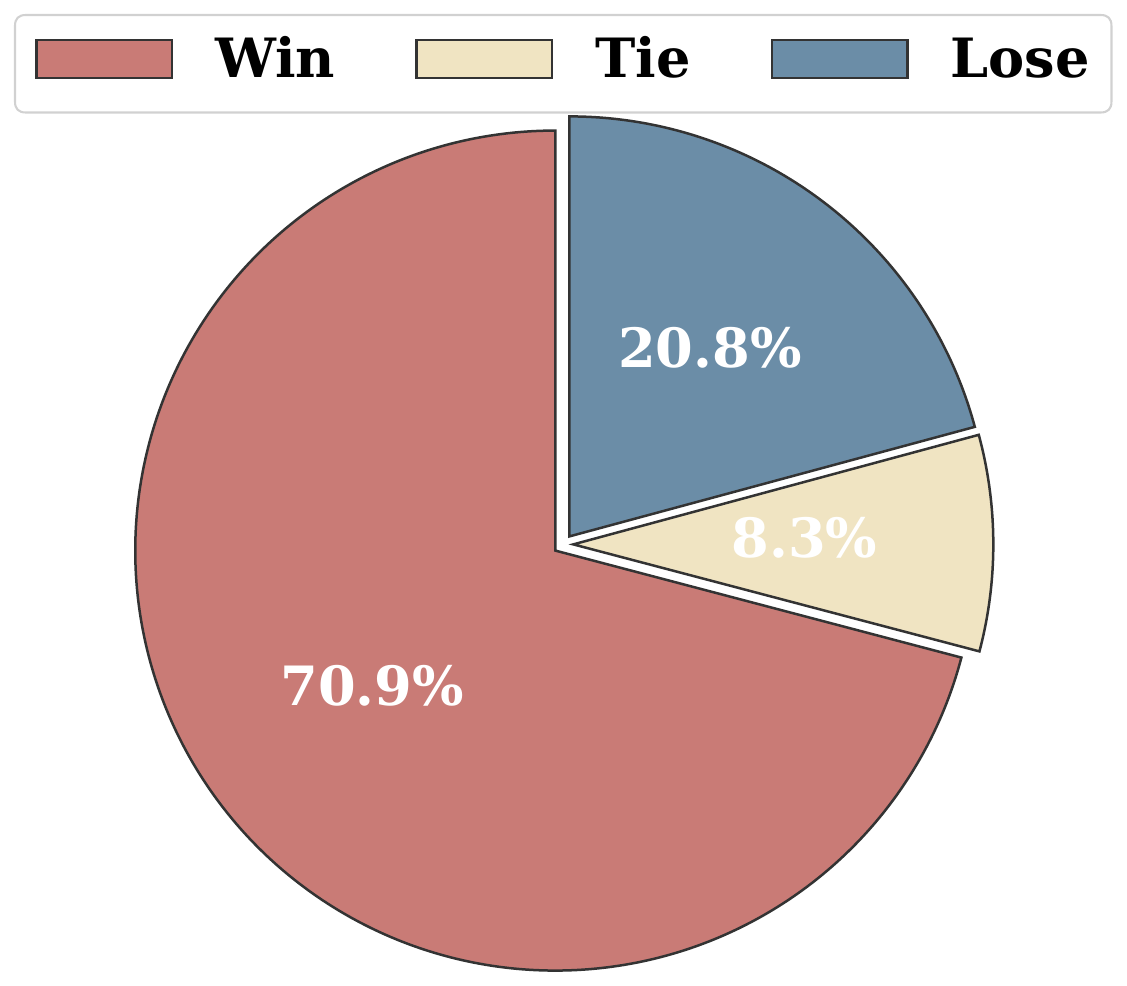}
    \label{fig:evolved_quality}
  \end{subfigure}
  \vspace{-2em}
 \caption{Effect of trace-driven skill evolution. Left: token usage and reasoning turns before and after skill evolution. Right: pairwise artifact quality comparison (Win/Tie/Lose) between the original and evolved skill libraries.}
  \label{fig:skill_evolution}
  \vspace{-2em}
\end{figure}

\stitle{Effect of skill evolution.}
We evaluate trace-driven skill evolution, split on offline production delivery cases into 50\% training cases for deriving skill refinements and 50\% held-out test cases for evaluation.
For each held-out case, we replay the same request with the original and evolved skill libraries, and measure token usage, reasoning turns, and artifact quality.
Artifact quality is assessed by pairwise LLM-as-judge comparison between artifacts generated before and after skill evolution, using position swapping and multi-model voting to reduce order bias and model-specific variance.
The detailed judging protocol provided in Appendix~\ref{app:skill_evolution_judge}.
As shown in Figure~\ref{fig:skill_evolution}, evolved skills reduce token usage by 25.5\% and reasoning turns by 17.5\% on held-out cases
For artifact quality, evolved artifacts are preferred or judged comparable in 79.2\% of cases.
These results suggest that skill evolution substantially reduces execution overhead without measurable degradation in final artifact quality.

\stitle{Effect of different LLM backbones.}
To examine whether the effectiveness of \method generalizes across
different LLM backbones, we conduct a paired evaluation using a
controlled agent runtime.
We adopt CodeBuddy as the agent runtime for this study, since it supports pluggable custom model backbones, whereas Codex and Claude Code do not expose such an interface to the best of our knowledge.
For each backbone, we compare two configurations: an agent equipped
with the shared warehouse skill library and the same agent augmented
with the three components of \method.
The backbone is fixed within each pair, while the runtime, tools,
prompts, context budget, refinement budget, and execution environment
are kept identical.
We evaluate representative enterprise-approved models spanning
different capability levels and model families.
As shown in Table~\ref{tab:paired-backbones}, enabling \method
consistently improves end-to-end delivery success across all evaluated
backbones.
The paired success gains range from 12.0 to 16.0 percentage points.
Notably, weaker backbones benefit more from \method (e.g., a 16.0-point gain on DeepSeek V4 Flash versus 12.0 points on the stronger GPT-5.5), as the deterministic platform evidence and structured orchestration compensate for limited reasoning capability.
Since all backbones are evaluated under the same CodeBuddy runtime, the reported numbers are not directly comparable to the product-conditioned results in Table~\ref{tab:offline-main}, which use each product's native runtime and backbone.
\method also reduces the number of reasoning turns, wall-clock time,
and token consumption for all backbones.
These results indicate that the effectiveness of \method does not
depend on a particular LLM backbone and that its structured delivery
workflow provides complementary benefits across different model
capability levels.

\stitle{Ablation studies.}
On the best-performing product, Claude Code, we remove one \method component at a time while holding the others fixed; Table~\ref{tab:offline-main} shows that all three components contribute, with the full model corresponding to the \textit{Claude Code + \method} row.
Removing hierarchical skill orchestration causes the largest degradation, an 11.5-point success drop (86.0\%$\rightarrow$74.5\%), concentrated on synchronization tasks where missing metadata, workflow context, or dependency evidence leads to invalid plans.
Removing artifact lifecycle control reduces success by 8.0 points (86.0\%$\rightarrow$78.0\%), mainly on computation tasks with latent correctness and performance risks; its pre- and post-execution sub-ablations both contribute to this drop.
Removing trace-driven skill evolution causes a relatively smaller 2.0-point success drop but increases total token consumption from 86.9K to 112.3K, showing that evolution primarily improves efficiency by compressing redundant context acquisition and intermediate reasoning.


\subsection{Online Deployment}
\label{sec:online_performance}

We deployed \method on Tencent Cloud WeData and evaluated it in a real production environment.
\method provides a web interface where engineers submit warehouse requirements, inspect generated artifacts, and track delivery progress; an example is shown in Figure~\ref{fig:method_UI} in Appendix~\ref{app:web-interface}.
The deployment covers 6 business teams and 4 warehouse task types: offline computation, offline synchronization, real-time computation, and real-time synchronization.
These tasks support core business scenarios such as WeChat Pay, advertising, and security risk control.

\begin{figure}
    \centering
    \includegraphics[width=0.99\linewidth]{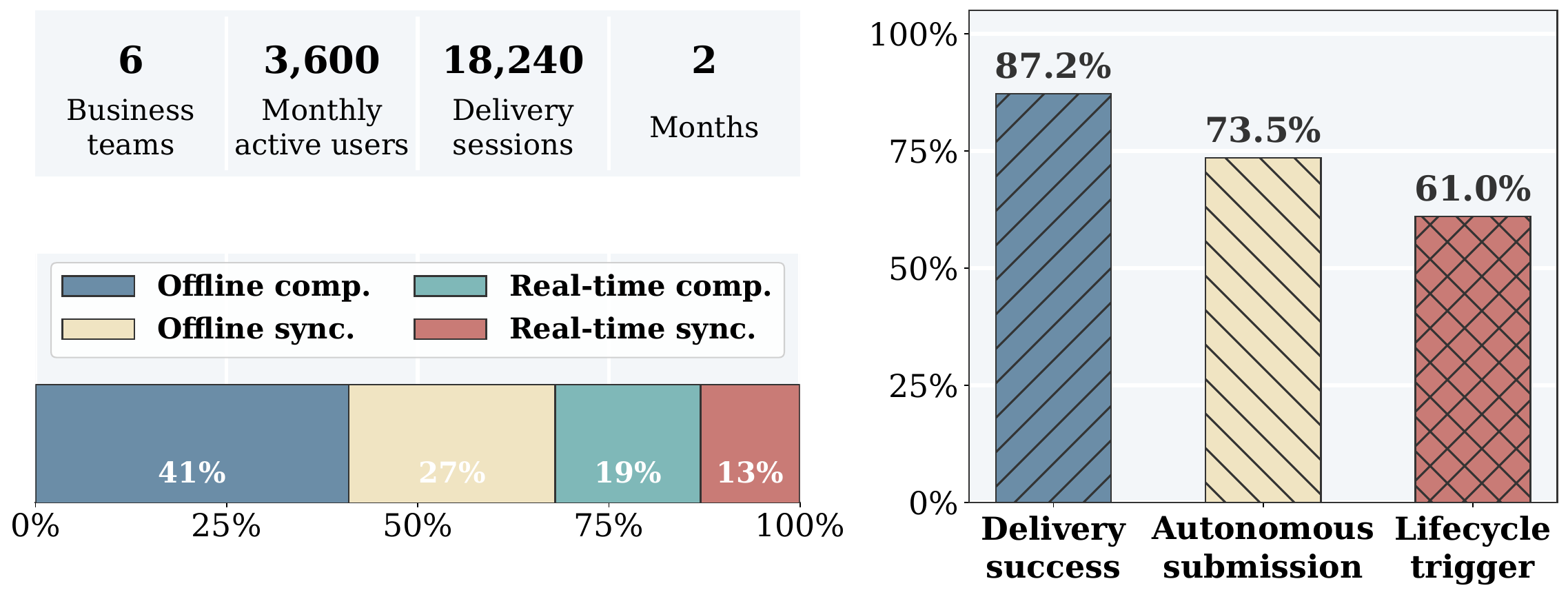}
    \vspace{-1em}
    \caption{Online production deployment of \method. Left: deployment scale and workload composition. Right: production delivery metrics.}
    \label{fig:online_results}
    \vspace{-1em}
\end{figure}

\stitle{Online results.}
As shown in Figure~\ref{fig:online_results}, during a two-month deployment period, \method served 3,600 monthly active users and supported 18,240 delivery sessions.
The four task types account for 41\%, 27\%, 19\%, and 13\% of online requests, respectively.
Across all sessions, \method achieved an end-to-end delivery success rate of 87.2\% and an autonomous submission rate of 73.5\%, where autonomous submission means that artifact generation, validation, and submission are completed without manual modification of delivery artifacts.
Artifact lifecycle control was triggered in 61\% of sessions: pre-execution diagnosis detected 4,120 risky artifacts before submission, and post-execution diagnosis repaired 2,890 platform execution failures.
Trace-driven skill evolution produced 128 human-approved skill updates, including 89 repair-oriented updates and 39 compression-oriented updates.
Based on 3,180 user ratings, \method received an average score of 4.4 out of 5.0.

\stitle{A/B test.}
We conducted a one-month A/B test on real online delivery requests.
We recruited 70 experienced data warehouse engineers who regularly perform warehouse task delivery and randomized them at the user level into a control group and a treatment group, with 35 users in each group.
The randomization was stratified by task type and business team.
The control group followed the existing manual workflow, while the treatment group used \method.
During the experiment, the treatment and control groups completed 718 and 682 delivery tasks, respectively.

\begin{figure}
    \centering
    \includegraphics[width=0.99\linewidth]{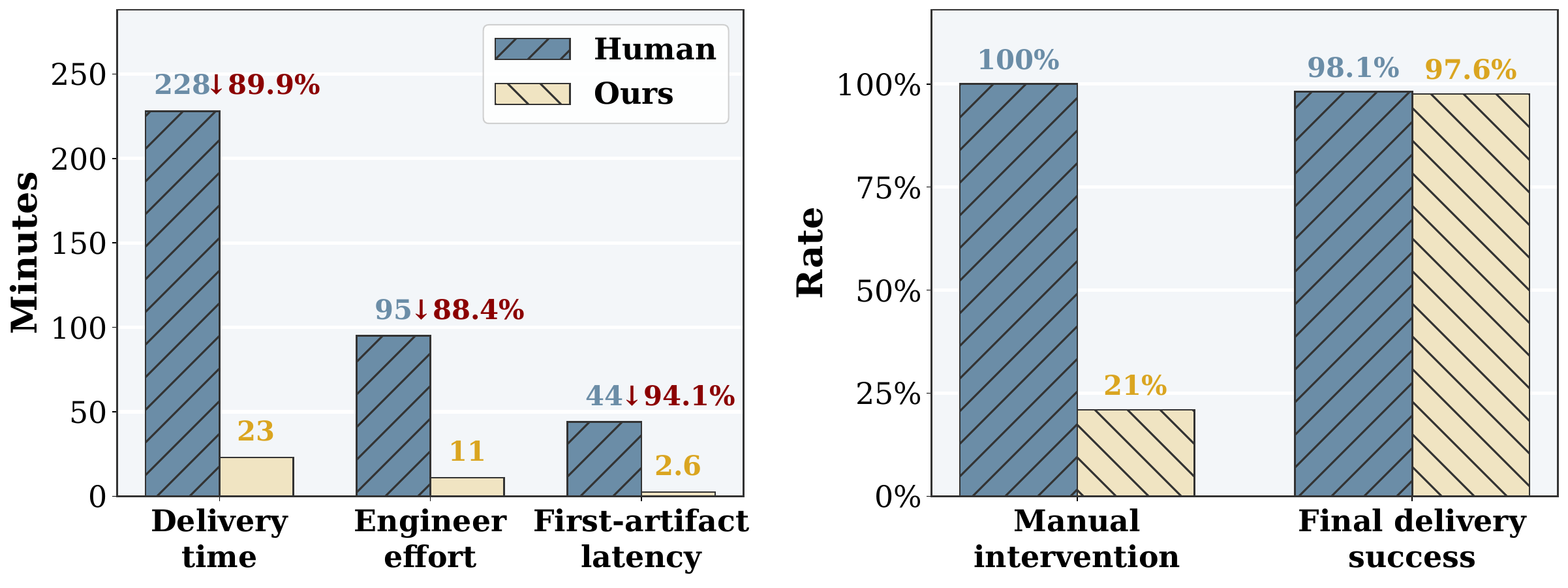}
    \vspace{-1em}
    \caption{Online A/B test comparing the manual workflow and \method. Left: delivery efficiency metrics. Right: human intervention and final delivery quality.}
    \label{fig:online_AB}
    \vspace{-1em}
\end{figure}

Since experienced engineers already achieve high final success rates, the A/B test focuses on delivery cost.
As shown in Figure~\ref{fig:online_AB}, compared with the control group, \method reduced the median end-to-end delivery time from 228 minutes to 23 minutes, corresponding to an 89.9\% reduction with a 95\% confidence interval of [86.4\%, 92.1\%].
It also reduced net engineer effort from 95 to 11 minutes, time to first executable artifact from 44 to 2.6 minutes, and manual intervention rate from 100\% to 21\%.
The treatment group required 3.2 interaction rounds on average, indicating that most sessions were completed without repeated clarification.
Meanwhile, final delivery success rates remained comparable between the treatment and control groups,  97.6 \% vs. 98.1\%, suggesting that efficiency gains did not compromise production submission quality.

\stitle{Production insights.}
The deployment yields three insights.
\textit{First}, gains in first-artifact latency, engineer effort, and manual intervention show that delivery automation should coordinate the  end-to-end delivery process rather than focus on code generation alone.
\textit{Second}, lifecycle control is essential for production readiness: its activation in 61\% of production sessions confirms the practical need to verify artifacts before submission and ground repairs in execution feedback.
\textit{Third}, the 128 approved updates show that warehouse skills are continually evolving operational assets; repair-oriented updates improve robustness, while compression-oriented updates reliably remove redundant context and interactions.


\stitle{Case studies.}
Appendix~\ref{sec:cases} presents four cases that complement the aggregate results with production traces.
In the pre-execution case, \method combines SQL semantics with table metadata to block an excessive partition scan after discovering approximately 1,179~TB in the first matching hourly partition.
In the post-execution case, it grounds the repair of a PySpark runtime failure in execution logs, artifact context, and retrieved knowledge, then generates a revised artifact for safe retry.
The skill-evolution case abstracts recurring empty-result and file-conflict failures into reusable preventive rules.
The end-to-end case follows an offline computation task from intent routing through artifact generation, validation, execution, and feedback-driven resubmission.
These cases demonstrate how \method integrates generation, lifecycle control, and skill evolution into a closed-loop production workflow.



\section{Conclusion}

We presented \method, an end-to-end production warehouse delivery system that goes beyond general coding agents through dependency-aware skill orchestration, lifecycle-aware artifact validation and repair, and trace-driven skill evolution.
Offline evaluations demonstrate improved delivery performance, effective diagnosis and repair, and reduced reasoning overhead.
A two-month deployment on Tencent Cloud WeData served 3,600 monthly active users and supported 18,240 delivery sessions, achieving an 87.2\% end-to-end success rate and a 73.5\% autonomous submission rate.
A production A/B test substantially reduced delivery time and engineer effort while maintaining comparable final delivery success, demonstrating that production warehouse automation requires coordinated planning, explicit artifact control, and experience-driven skill maintenance beyond code generation.


\clearpage
\bibliographystyle{ACM-Reference-Format}
\bibliography{reference}

\clearpage
\appendix
\section{Experimental Details}

\subsection{Detailed Evaluation Controls}
\label{app:evaluation-controls}

Within each product pair, both configurations are initialized from the same base warehouse skill library.
Agent states and temporary execution contexts are reset between cases, and no test-case outcome or platform feedback is used to update the skills or execution policy of subsequent cases.
Historical trajectories used for skill evolution are disjoint from evaluation cases, and the resulting skill-library snapshot is frozen before evaluation.

For the backbone study, all LLMs are accessed through direct APIs and integrated into the same CodeBuddy runtime on the same 200 cases.
The prompt template, skill-library snapshot, tool access, context budget, refinement budget, and execution environment are fixed within each pair; only the underlying model differs.
Backbone selection criteria are described in Section~\ref{sec:exp-offline}.

\subsection{Private Benchmark Construction and Public Companion Benchmark}
\label{app:benchmark}

\stitle{Case collection and stratification.}
The private benchmark was constructed from production warehouse delivery requests collected from Tencent Cloud WeData during the first half of 2026.
The unit of evaluation is one delivery request together with the platform context required to complete it, including the relevant table schemas, workflow metadata, task configurations, historical references, and execution permissions.
We retained requests for which the required platform context and the final delivery outcome were available, and excluded incomplete, cancelled, and duplicated requests.
Eligible requests were grouped into four task types: real-time synchronization, offline synchronization, real-time computation, and offline computation.
For each task type, we applied stratified selection and manually curated 50 representative cases, yielding 200 cases in total.
To reduce selection bias, the curated cases were chosen to span different difficulty levels, artifact types (e.g., SQL, PySpark, and Flink SQL), and failure modes, rather than favoring easily solved requests.
The same fixed cases and platform-context snapshots were used by all compared methods throughout all evaluations.

\stitle{Deduplication and leakage control.}
Before evaluation, we removed exact-duplicate delivery requests so that no request appears more than once in the benchmark.
The historical delivery trajectories used for trace-driven skill evolution are disjoint from the 200 evaluation cases.
The skill-library snapshot, task inputs, metadata, and platform context were frozen before evaluation, and no evaluation outcome or platform feedback was used to update the skills or execution policy of subsequent cases.

\stitle{Success adjudication.}
A case is counted as successful only when the generated artifact bundle is accepted by the platform, is executable in the designated environment, and satisfies the required artifact, dependency, and configuration constraints.
The final outputs were independently verified by three experienced warehouse engineers using a predefined checklist, and disagreements were resolved through adjudication.
For each task type and configuration, we report integer success counts together with the corresponding success rates for cross-method performance comparison.

\stitle{Public companion benchmark.}
The private benchmark cannot be publicly released because the cases contain proprietary schemas, workflow configurations, platform metadata, operational traces, and business-sensitive information.
To support public research on related data-engineering agent capabilities, we separately provide DataClawEval, a public companion benchmark of 100 executable tasks across PySpark, MySQL, HiveSQL, PrestoSQL, and Flink SQL.
Each task provides a sanitized request, an initialized data environment, a ground-truth solution, an automated grader, and a Docker-based execution harness.
DataClawEval complements rather than reproduces the private benchmark: it covers offline and real-time computation tasks, but does not model WeData-specific synchronization, scheduling, permission, platform-submission, or lifecycle-diagnosis workflows.
The DataClawEval tasks do not contain any of the 200 private evaluation cases and are not used to produce the results reported in this paper under any setting.

\subsection{LLM-as-Judge Protocol for Skill Evolution}
\label{app:skill_evolution_judge}
We use a pairwise LLM-as-judge protocol to assess whether skill evolution affects artifact quality.
For each case, we compare two artifacts generated from the same request: one using the original skill library and the other using the evolved skill library.
The judge is asked to select the better artifact according to functional correctness, dependency consistency, configuration validity, and production readiness.
To reduce position bias, each case is evaluated twice by the same judge with swapped artifact order.
If the two judgments are consistent, we record the result as a win for the corresponding artifact.
For example, if the judge selects artifact $A$ in both the original order and the swapped order, the case is recorded as an $A$ win for this judge.
If the two judgments are inconsistent, the case is recorded as inconsistent for subsequent analysis.

We use three LLM judges: Gemini 3.8 Flash, GPT-5.5, and Claude Opus 4.8.
Each judge independently produces one of three outcomes for each case: $A$ win, $B$ win, or inconsistent.
We then aggregate the three judge outcomes by majority voting.
If two or more judges select the same artifact, the result is recorded as a win for that artifact.
For example, outcomes $(A, A, B)$ are recorded as an $A$ win.
If no artifact receives at least two votes, or if the votes are dominated by inconsistent outcomes, the result is recorded as inconsistent.
For example, outcomes $(A, B, \text{inconsistent})$ are recorded as inconsistent.

\begin{table*}[!t]
\centering
\small
\caption{Diagnosis coverage in \method. Pre-execution diagnosis detects risks before task submission, while post-execution diagnosis structures feedback from platform logs.}
\label{tab:diag_coverage}
\begin{tabular}{l|lll}
\toprule
\rowcolor{gray!6}
Stage & Issue type & Main signal & Risk or feedback \\
\midrule
\multirow{8}{*}{Pre-execution}
 & Syntax error & SQL parser or compiler evidence & Invalid SQL or dialect-specific misuse \\
& Brute-force scan & Missing or ineffective partition pruning & Excessive scan cost and task delay \\
 & Cartesian product & Missing or incomplete join conditions & Data explosion and resource waste \\
& Implicit conversion & Type mismatch in filters, joins, or expressions & Semantic drift or pruning failure \\
& Empty-table read & Empty source table or missing partition & Silent empty outputs or zero metrics \\
 & Basic logic defect & Aggregation, join, null, or dedup logic & Silent data distortion \\
 & Where-clause logic defect & Boundary, null, or predicate logic & Incorrect filtering results \\
 & Business-intent mismatch & SQL semantics vs. user intent & Correct execution but wrong business meaning \\
\midrule
\multirow{6}{*}{Post-execution}
& Syntax exception & Parse, alias, function, join, or subquery errors & Root cause for failed compilation/execution \\
& UDF exception & UDF runtime, dependency, or registration errors & UDF repair or dependency fix \\
& Permission exception & Access-denied or unauthorized logs & Permission application or policy check \\
& System environment exception & Timeout, OOM, network, or service failures & Runtime or resource-level repair \\
& Data-source read exception & Missing file, corrupted data, or source failure & Source availability or format repair \\
 & Table-structure exception & Missing table, column, partition, or schema mismatch & Metadata or schema repair \\
\bottomrule
\end{tabular}
\end{table*}

\section{Case Studies}
\label{sec:cases}

\subsection{Case Study of Pre-execution Diagnosis}
\label{app:pre_diagnosis_case}
\begin{figure}
    \centering
    \includegraphics[width=0.99\linewidth]{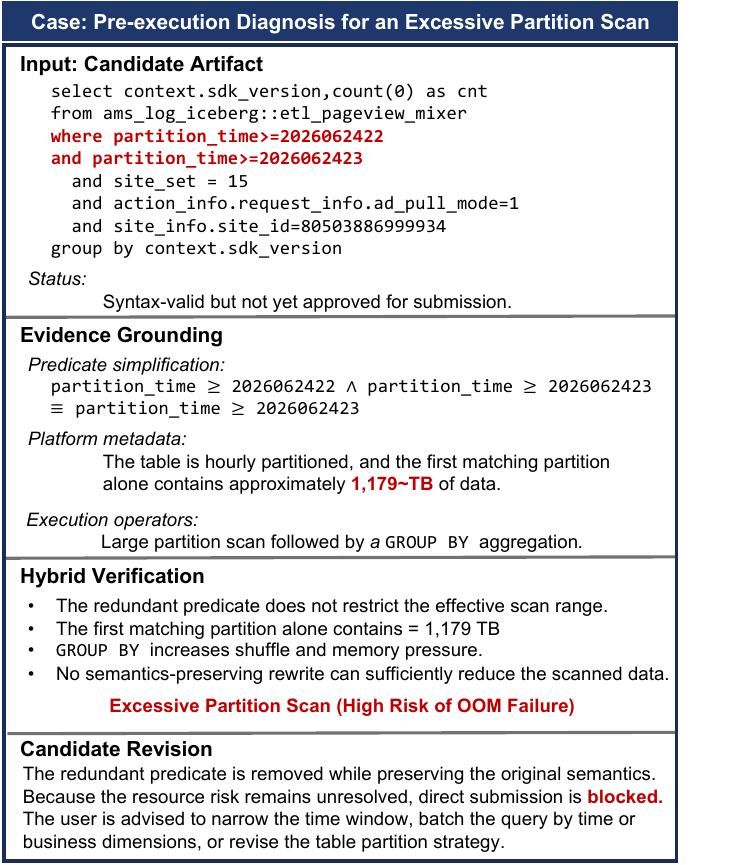}
    \vspace{-1em}
\caption{Case study of pre-execution diagnosis for an excessive partition scan. By combining SQL predicate analysis with table metadata, \method identifies a high risk of resource exhaustion, blocks direct submission, and returns a semantics-preserving revision with mitigation guidance.}
    \label{fig:pre-diagnosis-case}
\end{figure}

Figure~\ref{fig:pre-diagnosis-case} presents a case in which a syntactically valid SQL query remains unsafe for production execution.
The query contains two overlapping lower-bound predicates on \texttt{partition_time}, whose conjunction is equivalent to \texttt{partition_time >= 2026062423}.
Predicate simplification alone does not reduce the effective scan range.
Pre-execution diagnosis further retrieves table metadata and finds that the first matching hourly partition alone contains approximately 1,179~TB of data.
Combined with the subsequent \texttt{GROUP BY} aggregation, this introduces a high risk of excessive shuffle, memory pressure, and OOM failure.
Since no semantics-preserving rewrite can sufficiently reduce the scanned data, lifecycle control removes the redundant predicate but does not approve the artifact for direct submission.
Instead, it recommends narrowing the time range, batching the query by time or business dimensions, or revising the table partition strategy.

This case illustrates that syntax validation alone is insufficient for production warehouse delivery.
Reliable diagnosis requires jointly reasoning over query semantics, table metadata, and execution characteristics, while avoiding unsupported rewrites when the underlying risk cannot be safely resolved.

\subsection{Case Study of Post-execution Diagnosis}
\begin{figure}
    \centering
    \includegraphics[width=0.99\linewidth]{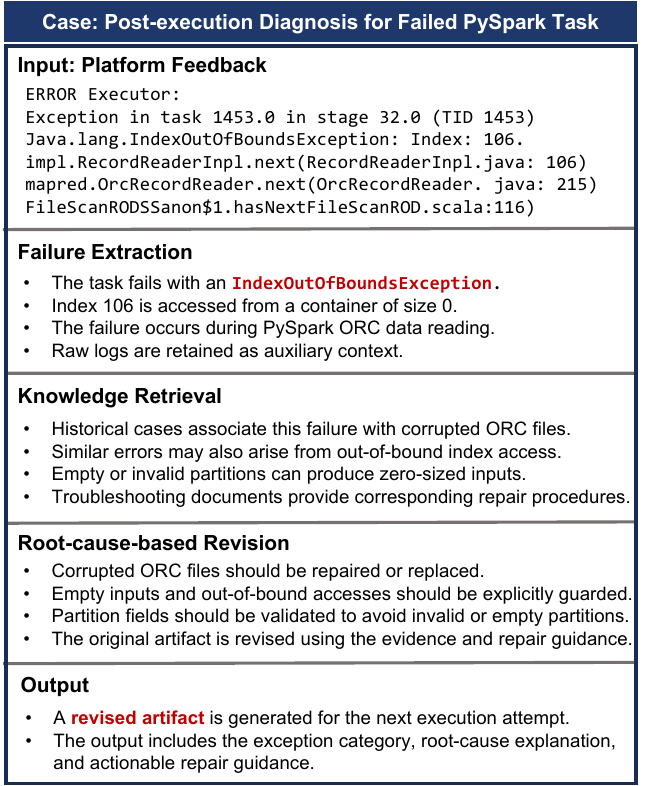}
    \vspace{-1em}
\caption{Case study of post-execution diagnosis for a PySpark runtime failure. By combining execution log analysis with diagnostic knowledge, \method identifies the root cause, provides actionable repair guidance, and generates a revised artifact for subsequent execution.}
    \label{fig:post-diagnosis-case}
    \vspace{-1em}
\end{figure}

Figure~\ref{fig:post-diagnosis-case} presents a case in which a PySpark task fails during runtime with an \texttt{IndexOutOfBoundsException}. The failure log reports an invalid access to index 106 from a container with size 0 during ORC data reading. Post-execution diagnosis first extracts localized failure signals from execution logs and retrieves relevant troubleshooting knowledge from historical cases. Based on the artifact context, failure signals, and retrieved knowledge, the diagnosis examines potential causes including corrupted ORC files, invalid reading logic, and empty input partitions.The evidence indicates that abnormal ORC data is the most likely root cause. The system then generates repair guidance to replace corrupted files, validate partition inputs, and add safeguards against empty data access. Using the localized evidence and repair guidance, an LLM revises the original artifact for the next execution attempt.

This case illustrates that execution logs alone are insufficient for reliable failure repair. Effective post-execution diagnosis requires combining platform feedback, artifact context, and historical knowledge to localize root causes and generate grounded revisions rather than directly modifying artifacts based on raw errors.

\begin{figure}
    \centering
    \includegraphics[width=0.99\linewidth]{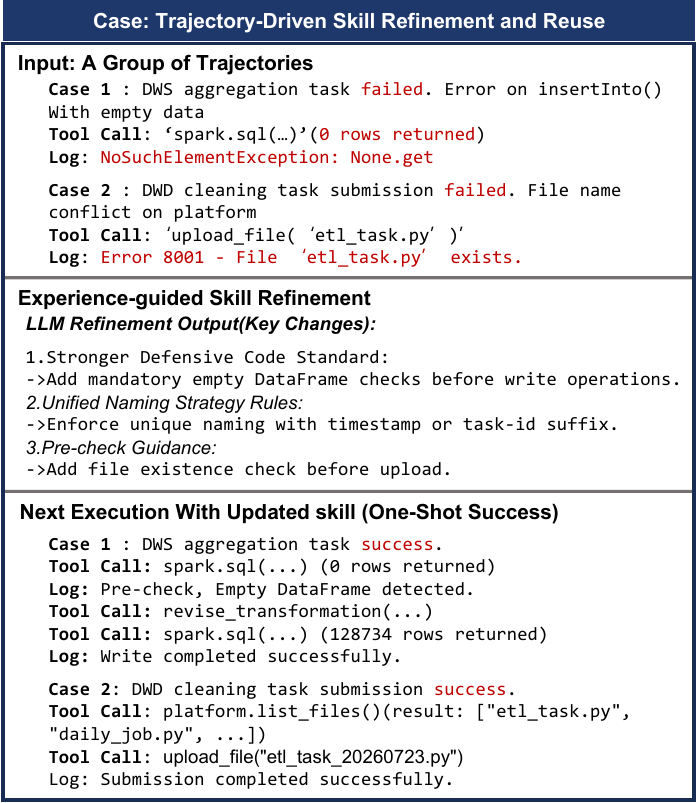}
    \vspace{-1em}
\caption{Case study of trajectory-driven skill refinement and reuse.}
    \label{fig:skill-evolved-case}
    \vspace{-1em}
\end{figure}

\subsection{Case Study of Trace-driven Skill Evolution}

Figure~\ref{fig:skill-evolved-case} presents a case in which execution traces are converted into reusable skill updates.
The trajectories contain two recurring failures: an empty DataFrame before \texttt{insertInto()} that later triggers a \texttt{NoSuchElementException}, and an \texttt{upload\_file('etl\_task.py')} operation that fails because the target file already exists.
Rather than treating them as isolated fixes, trace-driven skill evolution abstracts them into auditable, reusable delivery rules across similar tasks, including empty-result checks before writes, unique file naming, and platform-side existence checks before upload.

When the evolved skill is reused, these failures are prevented before platform submission.
The agent detects empty intermediate results, revises the transformation before writing, and selects non-conflicting filenames after checking existing platform files.
This case illustrates how \method turns failed trajectories into preventive execution knowledge, reducing repeated delivery failures through reusable skill refinement.

\begin{figure*}
    \centering
    \includegraphics[width=0.99\linewidth]{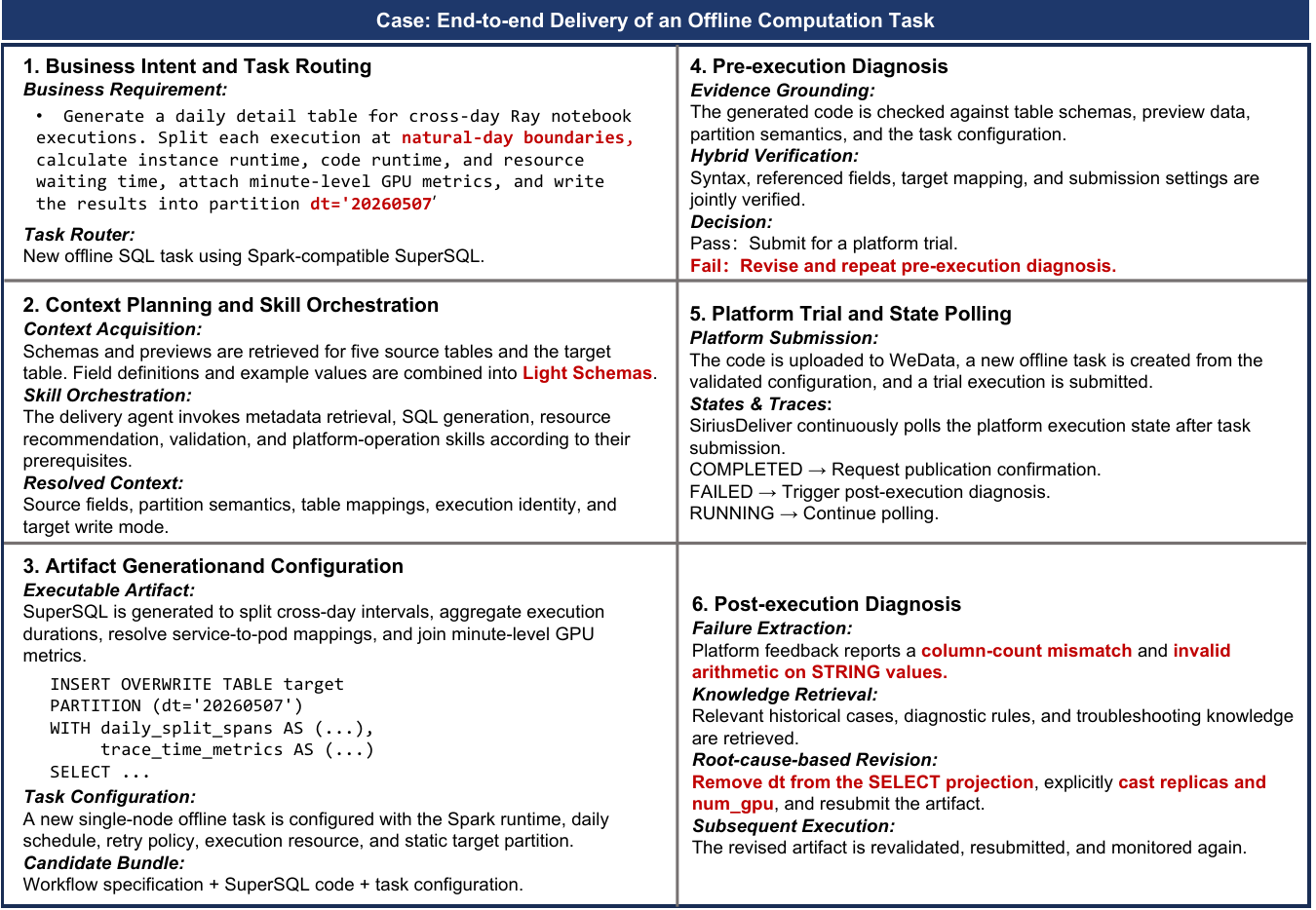}
    \vspace{-1em}
\caption{Case study of end-to-end delivery of an offline computation task.}
    \label{fig:e2e-case}
\end{figure*}

\subsection{Case Study of End-to-end Delivery}

Figure~\ref{fig:e2e-case} presents an end-to-end delivery case for an offline computation task.
The user requests a daily detail table for cross-day Ray notebook executions, requiring day-boundary splitting, runtime calculation, minute-level GPU metrics, and insertion into the target partition \texttt{dt='20260507'}.
The router identifies the request as a new offline SQL task, selects Spark-compatible SuperSQL, and orchestrates schema retrieval, preview-data acquisition, SQL generation, validation, resource recommendation, and platform operations.

Based on the resolved context, \method generates a workflow specification, SuperSQL code, and task configuration.
Pre-execution diagnosis checks schema consistency, partition semantics, referenced fields, target mappings, and submission settings before the artifact is uploaded to WeData for trial execution.
When the trial exposes a column-count mismatch and invalid arithmetic over \texttt{STRING} values, post-execution diagnosis removes the partition column from the \texttt{SELECT} projection, adds explicit casts, and resubmits the revised artifact.
This case illustrates how \method connects intent routing, context planning, artifact generation, lifecycle diagnosis, platform execution, and feedback-driven repair into a single production delivery loop.

\section{Details}

\subsection{Agent and skills}
The delivery automation agent operates over a reusable library of warehouse skills. A skill is the basic executable unit for warehouse delivery, consisting of
(1) a description of its functionality and applicable scenarios;
(2) a specification of input arguments extracted from the user request or context;
(3) execution code, which may involve LLM calls, platform APIs, retrieval modules, or deterministic scripts; and
(4) optional dependencies that determine execution order.
A skill may emit structured planning hints (e.g., inferred scenarios, remaining prerequisites, or downstream candidates)  to support subsequent planning.

\subsection{Web Interface}
\label{app:web-interface}

\begin{figure*}[t]
    \centering
    \includegraphics[width=0.85\textwidth]{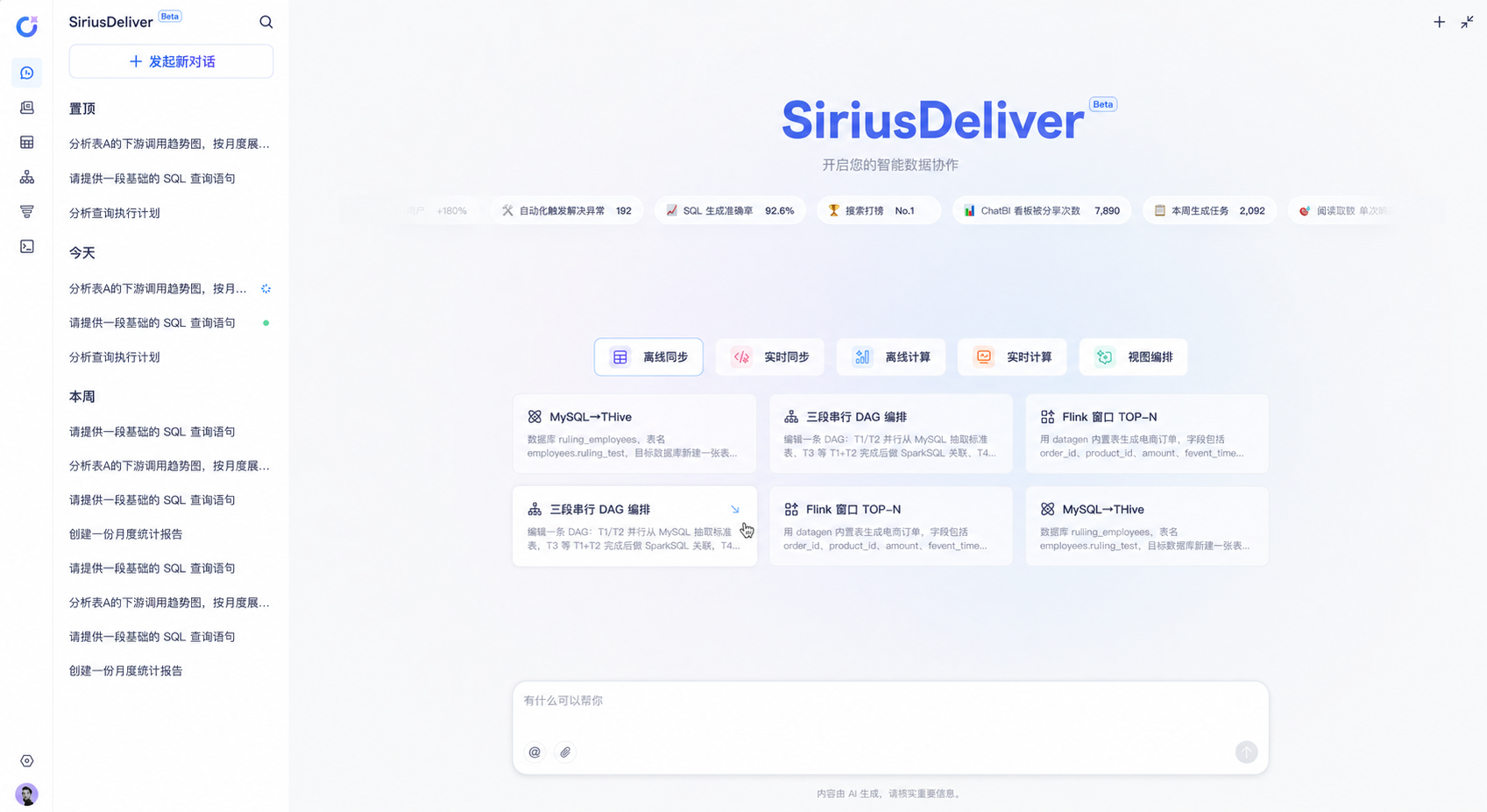}
    \vspace{-1em}
    \caption{Production web interface of \method.}
    \label{fig:method_UI}
\end{figure*}

Figure~\ref{fig:method_UI} shows the production interface of \method, where engineers submit warehouse requirements, inspect generated artifacts, and track delivery progress.
The interface provides a unified workspace for initiating new delivery conversations, revisiting historical sessions, and selecting common warehouse task scenarios.
Within each session, engineers can review the generated SQL, workflow configurations, and diagnostic feedback, provide clarifications when required information is missing, and monitor the status of validation, submission, and subsequent execution.
This interface operationalizes the human-in-the-loop delivery workflow by keeping requirement interaction, artifact inspection, and platform feedback within a single production-facing entry point.

\end{document}